\documentstyle[12pt]{article}
\newlength{\extraspace}
\newlength{\extraspaces}
\newcommand{\newsection}[1]{
\vspace{7mm}
\pagebreak[3]
\addtocounter{section}{1}
\setcounter{equation}{0}
\setcounter{subsection}{0}
\setcounter{footnote}{0}
{\large {\bf \thesection. #1}}
\nopagebreak
\medskip
\nopagebreak
\hspace{3mm}}

\newcommand{\ba}{\begin{eqnarray}
\addtolength{\abovedisplayskip}{\extraspaces}
\addtolength{\belowdisplayskip}{\extraspaces}
\addtolength{\abovedisplayshortskip}{\extraspace}
\addtolength{\belowdisplayshortskip}{\extraspace}}
\newcommand{\ea}{\end{eqnarray}}

\newcommand{\nonu}{\nonumber \\[.5mm]}
\newcommand{\A}{&\!\!\!}

\begin{document}
\thispagestyle{empty}

\begin{flushright}
SIT-LP-03/06 \\
{\tt hep-th/0306240} \\
June, 2003
\end{flushright}
\vspace{7mm}

\begin{center}
{\bf Toward linearization of nonlinear supersymmetric   \\
Einstein-Hilbert-type action of superon-graviton model(SGM) } \\[20mm]
{Kazunari SHIMA and Motomu TSUDA} \\[2mm]
{\em Laboratory of Physics, Saitama Institute of Technology}
\footnote{e-mail:shima@sit.ac.jp, tsuda@sit.ac.jp}\\
{\em Okabe-machi, Saitama 369-0293, Japan}\\[2mm]
{Manabu SAWAGUCHI} \\[2mm]
{\em High-Tech Research Center, Saitama Institute of Technology}
\footnote{e-mail:sawa@sit.ac.jp}\\
{\em Okabe-machi, Saitama 369-0293, Japan}\\[15mm]


\begin{abstract}
We attempt the linearization of the nonlinear supersymmetry 
encoded in the Einstein-Hilbert-type action 
${\it superon}$-${graviton \ model}(SGM)$ 
describing the nonlinear supersymmetric gravitational interaction 
of Nambu-Goldstone fermion ${\it superon}$. 
We discuss the linearizaton procedure in detail 
by the heuristic arguments referring to supergravity, 
in which particular attentions are paid to the local Lorentz 
invariance in the minimal interaction. 
We show explicitly up to the leading order that 80(bosons)+80(fermions) 
may be the minimal off-shell supermultiplet of the linearized theory.  
\end{abstract}
\end{center}

\newpage
\newsection{Introduction}

\noindent
In the previous paper\cite{ks1} we have proposed 
a new Einstein-Hilbert(E-H)-type superon-graviton model(SGM) 
action describing the nonlinear supersymmetrically(NLSUSY) \cite{va} 
invariant gravitational interaction of Nambu-Goldstone(N-G) 
fermion $superon$ in Riemann spacetime. 
SGM is obtained by extending the geometrical arguments of Einstein general 
relativity theory(EGRT) 
in Riemann spacetime to new (SGM) spacetime, where in addition to the 
ordinary Minkowski coordinates  the coset space coordinates 
of ${{N=10 \ superGL(4,R)} \over GL(4,R)}$ 
turning to the N-G fermion degrees of freedom(d.o.f.) 
are attached at every spacetime point\cite{ks1}.  
We have shown group theoretically that the SGM action 
with global SO(10) symmetry 
decomposed as ${\bf 10 = 5 + 5^{*}}$ with respect to SU(5) 
may give a unified description of spacetime and matter 
and proposed $superon$(-$quintet$)-$graviton$ $model(SGM)$\cite{ks2}. 
In SGM scenario all observed elementary particles are accomodated 
in a single irreducible representation 
of SO(10) super-Poincar\'e(SP) algebra. 
And except graviton they are regarded as the superon-composite 
eigenstates described in the low energy by the local fields 
of the ${linear}$ representation of supersymmetry(SUSY)\cite{wzgl} 
with (broken) SO(10) SP symmetry\cite{ks2} 
corresponding to the observed (low energy) nature.   
Considering that nature has $1 \times 2 \times 3$ gauge symmetry 
and SUSY GUTs usually contain 
more than 160 so many particles, we are tempted to suspect 
the elementariness and to imagine the specific 
internal structure  of spacetime and/or the compositeness 
of these (observed) particles. 
Due to the high nonlinear (self) couplings of the superon fields, 
as depicted in \cite{st1}, 
it is inevitble to linearize SGM for obtaining 
the equivalent low energy effective theory.  \\
In this work we would like to perform explicitly the linearization 
of N=1 SGM action of E-H-type 
to obtain the equivalent linear supersymmetric(LSUSY)\cite{wzgl} 
theory in the low energy, which is renormalizable. \\
Considering the abovementioned phenomenological potential of SGM, 
though qualitative and group theoretical so far, 
and the recent interest in NLSUSY in superstring(membrane) world, 
the linearization of NLSUSY in curved spacetime may be 
of some general interest. 

This paper is organized as follows. In Sec. 2 we briefly review 
the E-H-type action of SGM and the symmetries of the action. 
In Sec. 3 by heuristic arguments 
we carry out the linearizaton of NLSUSY in SGM referring 
the supermultiplet of $N = 1$ SUGRA 
and we show explicitly at the leading order that 
80(bosons)+80(fermions) may be the minimal off-shell supermultiplet 
of the linearized theory. 
The general discussion of the linearization procedure 
and the algebraic aspect is also described in Sec. 3. 
The conclusions are given in Sec. 4.

\newsection{New Einstein-Hilbert-type action of SGM}

\noindent
In this section, for the self-contained arguments 
we review N=1 SGM action briefly. 
Extending the geometrical arguments of EGRT on Riemann spacetime 
to new (SGM) spacetime where besides the Minkowski coordinate $x^{a}$ 
the coset space coordinates $\psi$ of ${superGL(4,R) \over GL(4,R)}$ 
turning to the N-G fermion d.o.f. are attached 
at every Riemann spacetime point, 
we obtain the following N=10 SGM action\cite{ks1}; 
\begin{equation}
L_{SGM}=-{c^{3} \over 16{\pi}G}\vert w \vert(\Omega + \Lambda ),
\label{SGM}
\end{equation}
\begin{equation}
\vert w \vert={\rm det}\ {w^{a}}_{\mu}
={\rm det}({e^{a}}_{\mu}+ {t^{a}}_{\mu}),  \quad
{t^{a}}_{\mu}={i \over 2} \kappa^4 \sum_{i=1}^{10}(\bar{\psi}^i \gamma^{a} 
\partial_{\mu} {\psi}^i
- \partial_{\mu} {\bar{\psi}^i} \gamma^{a} {\psi}^i),
\label{w}
\end{equation} 
where $w^{a}{_\mu}$ and $e^{a}{_\mu}$ 
are the vierbeins of unified SGM spacetime 
\footnote{
The SGM spacetime\cite{ks1,st1}, on which the unified vierbein 
$w{^a}_\mu$ is defined, is the curved spacetime 
whose tangent spacetime have the d.o.f. of $\psi$ as 
{\it a Grasmann coordinate} of the basic manifold 
besides the Minkowski coordinates $x^a$ 
(i.e., SO(3,1) $\times$ SL(2,C) d.o.f.). 
The unified vierbein $w{^a}_\mu$ 
is defined through $\omega^a = d x^a + i \kappa^4 \bar\psi \gamma^a d \psi 
= w^a{}_\mu d x^\mu$, where $\omega^a$ is the NLSUSY invarinat 
differential one-form of Volkov-Akulov(V-A)\cite{va}. 
}
and Riemann spacetime of EGRT 
respectively, $\psi^i$ ($i,j, ... =1,2, ... ,10$) is N-G fermion(superon) 
originating from the coset space coordinates 
of ${N=10 \ superGL(4,R) \over GL(4,R)}$, 
G is the gravitational constant, 
${\kappa^{4} = ({c^{3}\Lambda \over 16{\pi}G}})^{-1} $ 
is a fundamental volume of 
four dimensional spacetime of Volkov-Akulov(V-A) model\cite{va},  
and $\Lambda$ is a  ${small}$ cosmological constant related 
to the strength of the superon-vacuum coupling constant. 
Therefore SGM contains two mass scales ${1 \over G}$(Planck scale) 
and $\kappa \sim {\Lambda \over G}(O(1))$. 
$\Omega$ is a new scalar curvature analogous to 
the Ricci scalar curvature $R$ of EGRT, 
whose explicit expression is obtained  by just replacing 
${e^{a}}_{\mu}(x)$ by ${w^{a}}_{\mu}(x)$ in Ricci scalar $R$\cite{st1}. 
The SGM action (\ref{SGM}) is invariant at least 
under the following symmetries\cite{st2}; 
global SO(10), ordinary local GL(4R), 
the following new NL SUSY transformation 
\begin{equation}
\delta^{NL} \psi^{i}(x) ={1 \over \kappa^{2}} \zeta^{i} + 
i \kappa^{2} (\bar{\zeta}^{j}{\gamma}^{\rho}\psi^{j}(x)) \partial_{\rho}\psi^{i}(x),
\quad
\delta^{NL} {e^{a}}_{\mu}(x) = i \kappa^{2} 
(\bar{\zeta}^i {\gamma}^{\rho} \psi^i(x)) \partial_{[\rho} {e^{a}}_{\mu]}(x), 
\label{newsusy}
\end{equation} 
where $\zeta^i$ is a constant spinor 
and $\partial_{[\rho} {e^{a}}_{\mu]}(x) = 
\partial_{\rho}{e^{a}}_{\mu}-\partial_{\mu}{e^{a}}_{\rho}$, \\
the following GL(4R) transformations due to (\ref{newsusy})  
\begin{equation}
\delta_{\zeta} {w^{a}}_{\mu} = \xi^{\nu} \partial_{\nu}{w^{a}}_{\mu} + \partial_{\mu} \xi^{\nu} {w^{a}}_{\nu}, 
\quad
\delta_{\zeta} s_{\mu\nu} = \xi^{\kappa} \partial_{\kappa}s_{\mu\nu} +  
\partial_{\mu} \xi^{\kappa} s_{\kappa\nu} 
+ \partial_{\nu} \xi^{\kappa} s_{\mu\kappa}, 
\label{newgl4r}
\end{equation} 
where $\xi^{\rho}=i \kappa^{2} \bar{\zeta}^i {\gamma}^{\rho} \psi^i(x)$ 
and $s_{\mu \nu} = w{^a}_\mu w_{a \nu}$, \\
and the following local Lorentz transformation on $w{^a}_{\mu}$ 
\begin{equation}
\delta_L w{^a}_{\mu}
= \epsilon{^a}_b w{^b}_{\mu}
\label{Lrw}
\end{equation}
with the local parameter
$\epsilon_{ab} = (1/2) \epsilon_{[ab]}(x)$    
or equivalently on  $\psi^i$ and $e{^a}_{\mu}$
\begin{equation}
\delta_L \psi^i(x) = - {i \over 2} \epsilon_{ab}
      \sigma^{ab} \psi^i,     \quad
\delta_L {e^{a}}_{\mu}(x) = \epsilon{^a}_b e{^b}_{\mu}
      + {\kappa^{4} \over 4} \varepsilon^{abcd}
      \bar{\psi}{^i} \gamma_5 \gamma_d \psi{^i}
      (\partial_{\mu} \epsilon_{bc}).
\label{newlorentz}
\end{equation}
The local Lorentz transformation forms a closed algebra, 
for example, on $e{^a}_{\mu}(x)$ 
\begin{equation}
[\delta_{L_{1}}, \delta_{L_{2}}] e{^a}_{\mu}
= \beta{^a}_b e{^b}_{\mu}
+ {\kappa^{4} \over 4} \varepsilon^{abcd} \bar{\psi}{^i}
\gamma_5 \gamma_d \psi{^i}
(\partial_{\mu} \beta_{bc}),
\label{comLr1/2}
\end{equation}
where $\beta_{ab}=-\beta_{ba}$ is defined by
$\beta_{ab} = \epsilon_{2ac}\epsilon{_1}{^c}_{b} -  \epsilon_{2bc}\epsilon{_1}{^c}_{a}$.
The commutators of two new NLSUSY transformations (\ref{newsusy}) 
on $\psi^j(x)$ and  ${e^{a}}_{\mu}(x)$ 
are GL(4R), i.e. new NLSUSY (\ref{newsusy}) is the square-root of GL(4R); 
\begin{equation}
[\delta_{\zeta_1}, \delta_{\zeta_2}] \psi^i 
= \Xi^{\mu} \partial_{\mu} \psi^i, 
\quad
[\delta_{\zeta_1}, \delta_{\zeta_2}] e{^a}_{\mu}
= \Xi^{\rho} \partial_{\rho} e{^a}_{\mu}
+ e{^a}_{\rho} \partial_{\mu} \Xi^{\rho},
\label{com1/2-e}
\end{equation}
where 
$\Xi^{\mu} = 2i\kappa (\bar\zeta^i_2 \gamma^\mu \zeta^i_1)
      - \xi_1^{\rho} \xi_2^{\sigma} e{_a}^{\mu}
      (\partial_{[\rho} e{^a}_{\sigma]})$.
They show the closure of the algebra. 
SGM action (\ref{SGM}) is invariant at least under\cite{st2}
\begin{equation}
[{\rm global\ NLSUSY}] \otimes [{\rm local\ GL(4,R)}] 
\otimes [{\rm local\ Lorentz}] 
\otimes [{\rm global\ SO(10)}],  \\
\label{sgmsymm}
\end{equation}
which is isomorphic to SO(10) SP whose single irreducible representation gives 
the group theoretical description of SGM\cite{ks2}.

Here we just mention the confusive local spinor transformation 
which leaves SGM action (\ref{SGM}) invariant. 
The following local spinor translation 
with a local parameter $\epsilon(x)$, 
$\delta \psi = \epsilon$, 
$\delta e{^a}{_\mu} = - i \kappa^4 
(\bar\epsilon \gamma^a \partial_\mu \psi 
+ \bar\psi \gamma^a \partial_\mu \epsilon)$, 
gives $\delta w{^a}_\mu = 0 = \delta w{_a}^\mu$. 
However, it should be noticed that this local spinor 
transformation cannot transform away the d.o.f. of $\psi$. 
Indeed, $\psi$ seems to be transformed away 
if we choose $\delta \psi = \epsilon = - \psi$, 
but it is restored precisely in the unified vierbein $w{^a}_\mu$ 
by simultaneously transforming $e{^a}_\mu$, 
i.e., $w(e, \psi) = w(e + \delta e, \psi + \delta \psi) 
= w(e + t, 0)$ as indicated by $\delta w{^a}_\mu = 0$. 
And also the above local spinor transformation is a fake gauge 
transformation in a sense that, in contrast with the local Lorentz 
transformation on the coordinates in general relativity, 
it cannot be eliminate the d.o.f. of $\psi$ 
since the unified vierbein $w{^a}_\mu = e{^a}_\mu + t{^a}_\mu$ 
is the only gauge field on SGM spacetime and contains 
only integer spin. 
This confusive situation comes from the new geometrical formulation 
of SGM on unfamiliar SGM spacetime, where besides the Minkowski 
coordinates $x^a$, $\psi$ is a Grassmann coordinate 
(i.e. the fundamental d.o.f.) defining the tangential spacetime 
with SO(3,1) $\times$ SL(2,C) d.o.f. inspired by NLSUSY, 
and the local spinor transformation ($\delta \psi = \epsilon(x)$) 
is just a coordinate transformation(redefinition) on SGM spacetime. 
These situation can be understood easily by observing 
that the unified vierbein $w{^a}_\mu = e{^a}_\mu + t{^a}_\mu$ 
is defined by $\omega^a = d x^a + i \kappa^4 \bar\psi \gamma^a d \psi 
= w^a{}_\mu d x^\mu$, where $\omega^a$ is the NLSUSY invarinat 
differential one-form of V-A\cite{va} 
and $(x^a, \psi)$ are coordinates 
specifying the (SGM) flat spacetime inspired by NLSUSY. \\
From these geometrical viewpoints (in SGM spacetime) 
we can understand that $\psi$ is a coordinate and would 
not be transformed away, 
and the initial SGM spacetime is preserved. 
Therefore the action (\ref{SGM}) is a nontrivial 
generalization of the E-H action. 
Eliminating $\psi$ by some arguments 
regarding the above local spinor translation as a gauge transformation 
leads to a different theory (ordinary E-H action) 
with a different vacuum (Minkowski flat spacetime), 
which is another from SGM scenario considering that 
the SGM spacetime is an ultimate physical entity. 

The linearization of such a theory with a high nonlinearity 
is an interesting and inevitable to obtain 
an equivalent local field theory which is renormalizable 
and describes the observed low energy (SM) physics.

\newsection{Linearization}

\noindent
The linearizations of NLSUSY in flat spacetime have been carried out 
by many authors. 
They have proved ${\it algebraicly}$ that N=1 V-A model is equivalent to  
N=1 scalar supermultiplet\cite{ik}\cite{r}\cite{uz} action ${\it or}$ 
N=1 axial vector gauge supermultiplet action of LSUSY\cite{ik}\cite{stt2}. 
We have also proved by the heuristic arguments that N=2  V-A model 
is equivalent to N=2 LSUSY model 
with SU(2) invariance\cite{stt1} and ${\it vector \ J^{P}=1^{-}}$ 
gauge supermultiplet action is obtained 
by the spontaneous breakdown $SU(2) \rightarrow U(1)$ 
in the linear representation. 
Interestingly  the SU(2) gauge structure of the electroweak standard model(SM) 
may be explained for the first time provided that the electroweak gauge bosons 
are  the composite-eigenstates of these (SGM) types, however, 
described by asymptotic ${\it local}$ fields in the low energy. 
And the absence of the low energy excited states of observed gauge bosons, quarks and  leptons 
can be explained despite the compositeness.    \\
These equivalent LSUSY actions possess in general Fayet-Iliopoulos\cite{fi} terms 
indicating the spontaneous breakdown of LSUSY and SU(2). 
These algebraic exact  results are favourable to the SGM scenario.  \\
In those works of the linearization it is important to find the SUSY invariant relations  
which express  the LSUSY supermultiplets in terms of  NL theories and reproduce 
the LSUSY transformations on the linearized supermultiplet under the NLSUSY 
transformation expressed by superon fields.   \\
The SUSY invariant relations of the global SUSY in flat spacetime can be obtained 
straightforwardly, 
for the supermultiplets structures are well understood and 
the algebraic SP structures of both L and NL theories  are the same.    
These algebraic exact  results are favourable to the SGM scenario.  
However, it is unsatisfactory that the linearized theory is the  free theory of the supermultiplet  
and that the existing superfield thechnique would transform V-A action into the free theory 
by the formulation.    \\
The situation  in SGM is rather different from the flat space case, for the supermultiplet structure of 
the linearized theory of SGM is unknown except it is expected to be a broken SUSY SUGRA-like theory 
containing graviton and a (massive) spin 3/2 field  and 
the algebraic structure (\ref{sgmsymm}) is changed into  SP.       \\
Especially for N=10 SGM, the supermultiplet of the linearized theory 
should contain (massive) fields with spin up to 3, which may be  beyond the straightforward 
application of the existing (superfield) formalism restricted to $N<9$(i.e. spin$\leq 2$). 
Therefore  by the heuristic arguments and by referring to SUGRA  
we discuss the linearization for the moment. We focus to N=1 for simplicity.       \par
Following SGM scenario, we assume that;   \\
(i) the linearized theory possesses the  spontaneously broken ${\it global}$ (at least) SUSY,     \\
(ii) graviton is an elementary field  at least in the leading order 
(not composite of superons corresponding to the vacuum of the Clifford algebra) 
in both NL and L theories  and      \\
(iii) the NLSUSY supermultiplet of SGM ($e{^a}_{\mu}(x)$, $\psi(x)$) 
should be connected to the LSUSY composite supermultiplet 
(${\tilde e}{^a}_{\mu}(e(x), \psi(x))$, 
${\tilde \lambda}_{\mu}(e(x), \psi(x))$) 
for the SUGRA-like linearized theory. \par
From these assumptions and  the arguments spread 
in the flat space cases we require that 
the SUGRA gauge transformation \cite{fvfdz}
with the global spinor parameter ${\zeta}$ 
should hold for the supermultiplet (${\tilde e}{^a}_{\mu}(e, \psi)$, 
${\tilde \lambda_{\mu}(e, \psi)}$) 
of the (SUGRA-like) linearized theory, i.e.,  \\
\begin{equation}
\delta {\tilde e}{^a}_{\mu}(e, \psi)  
      = i\kappa \bar{ \zeta} \gamma^{a} {\tilde \lambda_{\mu}(e, \psi)}, 
\label{sugral-2}
\end{equation} 
\begin{equation}
\delta {\tilde \lambda}_{\mu}(e, \psi)  
      = {2 \over \kappa}D_{\mu}{ \zeta}  
      = -{i \over \kappa}\tilde \omega^{ab}{_\mu}\sigma_{ab}{ \zeta}, 
\label{sugral-3/2} 
\end{equation} 
where  $\sigma^{ab} = {i \over 4}[\gamma^a, \gamma^b]$, 
$D_{\mu}=\partial_{\mu}-{i \over 2} {{\omega}^{ab}{_\mu}}(e, \psi)\sigma_{ab}$, and ${ \zeta}$ is a 
global spinor parameter and the variations in the left-hand side are computed under NLSUSY (\ref{newsusy}). \par
${\bf 3.1\ Case \ \tilde e{^a}_{\mu}(e, \psi) = e{^a}_{\mu}}$ \\
We put the following  SUSY invariant relations which connect $e^{a}{_\mu}$ to ${\tilde e}{^a}_{\mu}(e, \psi)$;
\begin{equation}
{\tilde e}{^a}_{\mu}(e, \psi) = { e}{^a}_{\mu}(x).     
\label{relation-2}
\end{equation} 
The relation (\ref{relation-2}) is the assumption (ii) and the metric conditions holds simply.
Consequently the following invariant relation  is obtained by substituting  (\ref{relation-2}) 
into (\ref{sugral-2}) and  computing the variations under (\ref{newsusy})\cite{sts}; 
\begin{equation}
{\tilde \lambda}_{\mu}(e, \psi)  
      = \kappa \gamma_{a} \gamma^{\rho} \psi(x) \partial_{[\rho} e{^a}_{\mu]}.    
\label{relation-3/2} 
\end{equation} 
(As discussed later these should may be considered as the leading order of the expansions in $\kappa$ of 
SUSY invariant relations. The expansions terminate with $(\psi)^{4}$.)
Now we see LSUSY transformation 
%
%
induced by (\ref{newsusy}) on the (composite) supermultiplet 
(${\tilde e}{^a}_{\mu}(e, \psi)$, ${\tilde \lambda}_{\mu}(e, \psi$)).    \\
The LSUSY transformation  on $\tilde e{^a}_{\mu}$ becomes  as follows. 
The left-hand side of (\ref{sugral-2}) gives
\begin{equation}
\delta {\tilde  e}{^a}_{\mu}(e, \psi) = \delta^{NL} {e^{a}}_{\mu}(x) 
= i \kappa^{2} (\bar{\zeta}{\gamma}^{\rho}\psi(x))\partial_{[\rho} {e^{a}}_{\mu]}(x). 
\label{susysgm-2} 
\end{equation} 
While substituting (\ref{relation-3/2}) into the righ-hand side of  (\ref{sugral-2}) we obtain 
\begin{equation}
i \kappa^{2} (\bar{\zeta}{\gamma}^{\rho}\psi(x))\partial_{[\rho} {e^{a}}_{\mu]}(x) + \cdots({\rm extra \ terms}).
\label{susysgm-3/2} 
\end{equation} 
These results show that  (\ref{relation-2}) and (\ref{relation-3/2}) are not  SUSY invariant relations 
and reproduce (\ref{sugral-2}) with unwanted extra terms which should be identified with the auxirialy 
fields. 
The commutator of the two LSUSY transformations induces GL(4R) with the field dependent parameters as follows;
\begin{equation}
[\delta_{\zeta_1}, \delta_{\zeta_2}]{\tilde  e}{^a}_{\mu}(e, \psi) 
= \Xi^{\rho} \partial_{\rho} {\tilde  e}{^a}_{\mu}(e, \psi)    
+ {\tilde  e}{^a}_{\rho}(e, \psi)\partial_{\mu} \Xi^{\rho},
\label{susysgmcom-2}
\end{equation}
where 
$\Xi^{\mu} = 2i (\bar{\zeta}_2 \gamma^{\mu} \zeta_1)
      - \xi_1^{\rho} \xi_2^{\sigma} e{_a}^{\mu}
      (\partial_{[\rho} e{^a}_{\sigma]})$.
On  ${\tilde \lambda}_{\mu}(e, \psi)$, the left-hand side of (\ref{sugral-3/2}) becomes 
apparently rather complicated; 
\ba
\delta {\tilde \lambda}_{\mu}(e, \psi)  
\A = \A {\kappa } \delta( \gamma_{a} \gamma^{\rho} \psi(x) \partial_{[\rho} e{^a}_{\mu]}) \nonu
\A = \A {\kappa } \gamma_{a}[ \delta^{NL}\gamma^{\rho} \psi(x) \partial_{[\rho} e{^a}_{\mu]} + 
  \gamma^{\rho} \delta^{NL} \psi(x) \partial_{[\rho} e{^a}_{\mu]} +
  \gamma^{\rho}  \psi(x) \partial_{[\rho} \delta^{NL} e{^a}_{\mu]}]. 
\label{susysgm-3/2} 
\ea
However the commutator of the two LSUSY transformations 
induces the similar GL(4R);
\begin{equation}
[\delta_{\zeta_1}, \delta_{\zeta_2}]{\tilde \lambda}_{\mu}(e, \psi)  
= \Xi^{\rho} \partial_{\rho} {\tilde \lambda}_{\mu}(e, \psi)  
+ {\tilde \lambda}_{\rho}(e, \psi)\partial_{\mu} \Xi^{\rho}.  
\label{susysgmcom-3/2}
\end{equation}                   \par
These results indicate that the algebra on the linearized field closes 
and the initial new NLSUSY 
structure of SGM is maintained on the linearized supermultiplet (,i.e. disappointedly the commutators does not induce SP symmetry) 
provided that the relations (\ref{relation-2}) and (\ref{relation-3/2}) 
and SUGRA transformation (\ref{sugral-2}) are respected.  
And due to the complicated expression of LSUSY (\ref{susysgm-3/2}) 
which makes the physical and 
mathematical structures are obscure, we can hardly guess a linearized invariant action 
which is equivalent to SGM.       \par
Now we attempt the linearization such that LSUSY transformation 
on the linearized fields induces SP transformation.   \par
By comparing (\ref{sugral-3/2}) with (\ref{susysgm-3/2}) we understand that the local Lorentz transformation  
plays a crucial role. 
As for the local Lorentz transformation on the linearized asymptotic fields corresponding to the observed 
particles (in the low energy), 
it is natural to take (irrespective of (\ref{newlorentz})) the following forms   \\ 
\begin{equation}
\delta_L \tilde \lambda_{\mu}(x) = - {i \over 2} { \epsilon}_{ab}
      \sigma^{ab} \tilde \lambda_{\mu}(x),     \quad
\delta_L \tilde { e^{a}}_{\mu}(x) = {\epsilon}{^a}_b \tilde e{^b}_{\mu}, 
\label{lorentz}
\end{equation}
where $\epsilon_{ab} = (1/2)\epsilon_{[ab]}(x)$ is a local parameter.   
The relation between (\ref{newlorentz})   and (\ref{lorentz}) , i.e. the Lorentz invariance encoded 
geometrically in  SGM space-time  and (\ref{lorentz}) of the Lorentz invariance 
defined on the (composite) asymptotic field in Riemann(Minkowski) space-time, 
is unknown. 
However this is particularly interesting, for in SGM  the local Lorentz transformations  (\ref{Lrw}) 
and (\ref{newlorentz}), 
i.e. the local Lorentz invariant gravitational interaction of superon, 
are introduced by the geometrical arguments in SGM spacetime\cite{st2} 
following EGRT.  
While in SUGRA the local Lorentz transformation invariance 
(\ref{lorentz}) is realized as usual by introducing minimally 
the Lorentz spin connection $\omega{_\mu}^{ab}$. 
And the LSUSY transformation is defined successfully 
by the (Lorentz) covariant derivative containing the spin connection 
$\tilde \omega^{ab}{_\mu}(e,\psi)$ as seen in (\ref{sugral-3/2}), 
which causes the super-Poincar\'e algebra on the commutator of SUSY and is convenient for 
constructing the invariant action. 
Therefore in the linearized (SUGRA-like) theory the local Lorentz  transformation  invariance  is expected 
to be realized as usual by defining (\ref{lorentz}) and introducing the Lorentz spin connection 
$\omega^{ab}{_\mu}$.
We investigate how the spin connection $\tilde\omega^{ab}{_\mu}(e,\psi)$ appears 
in the linearized (SUGRA-like) theory through  the linearization process. 
This is also crucial for constructing a nontrivial (interacting) linearized action 
which has manifest invariances.  \par  
We discuss the Lorentz covariance of the  transformation by comparing (\ref{susysgm-3/2}) with 
the right-hand side of (\ref{sugral-3/2}). 
The direct computation of (\ref{sugral-3/2}) by using SUSY invariant relations (\ref{relation-2}) 
and (\ref{relation-3/2}) under (\ref{newsusy}) produces complicated redundant terms 
as read off from (\ref{susysgm-3/2}).
The local Lorentz invariance of the linearized theory may become ambiguous and lose the manifest invariance. \\
For a simple restoration of the manifest local Lorentz invariance 
we survey the possibility that such redundant terms may be recasted  by  the d.o.f of 
the auxiliary fields in the linearized supermultiplet.
As for the auxiliary fields it is necessary for the closure of the off-shell superalgebra 
to include the equal number of the fermionic and the bosonic d.o.f. in the linearized supermultiplet. 
As new NLSUSY is a global symmetry, 
${\tilde \lambda}_{\mu}$ has 16 fermionic d.o.f.. 
Therefore at least 4 bosonic d.o.f. must be added to the off-shell SUGRA supermultiplet 
with 12 d.o.f.\cite{swfv} and a vector field  may be a simple candidate.  
However, counting the bosonic d.o.f. present in the redundant terms corresponding to 
$\tilde \omega^{ab}{_\mu}(e,\psi)$, 
we may need a bigger supermultiplet  e.g. $16 + 4 \cdot 16 = 80$  d.o.f., to carry out the linearization.     \\ 
Now we consider  the  simple modification of SUGRA transformations(algebra) by adjusting 
the (composite) structure of the (auxiliary) fields.  
We take, in stead of (\ref{sugral-2}) and  (\ref{sugral-3/2}), 
\begin{equation}
\delta {\tilde e}{^a}_{\mu}(e, \psi) 
      = i\kappa \bar{\zeta} \gamma^{a} {{\tilde \lambda}_{\mu}(x)} + \bar{\zeta}{\tilde \Lambda}{^a}_{\mu}, 
\label{newsugral-2}
\end{equation} 
\begin{equation}
\delta {\tilde \lambda}_{\mu}(e, \psi) 
      = {2 \over \kappa}D_{\mu}\zeta + {\tilde \Phi}_{\mu}\zeta  
      = -{i \over \kappa}\tilde \omega^{ab}{_\mu}\sigma_{ab}\zeta + {\tilde \Phi}_{\mu}\zeta, 
\label{newsugral-3/2} 
\end{equation} 
where  $\tilde \Lambda{^a}_{\mu}$ and  $\tilde \Phi_{\mu}$ represent the auxiliary fields to be specified 
and are functionals of $e^{a}{_\mu}$ and $\psi$. 
We need $\tilde \Lambda{^a}_{\mu}$ term in (\ref{newsugral-2}) to alter (\ref{susysgm-2}), 
(\ref{susysgmcom-2}), (\ref{susysgm-3/2}) and  (\ref{susysgmcom-3/2}) 
toward that of super-Poincar\'e algebra of SUGRA.  
We attempt the restoration of the manifest local Lorentz invariance order by order by adjusting 
$\tilde \Lambda{^a}_{\mu}$ and  $\tilde \Phi_{\mu}$. 
In fact, the Lorentz spin connection  $\omega^{ab}{_\mu}(e)$(i.e. the leading order terms of 
$\tilde \omega^{ab}{_\mu}(e, \psi)$) of (\ref{newsugral-3/2}) is reproduced by taking the following one  
\begin{equation}
\tilde \Lambda{^a}_{\mu} = {\kappa^{2} \over 4}[ ie_{b}{^\rho}\partial_{[\rho}e{^b}_{\mu]}\gamma{^a}\psi 
- \partial_{[\rho}e_{\mid b \mid \sigma]}e{^b}_{\mu}\gamma^{a}\sigma^{\rho\sigma}\psi ], 
\label{auxlambda-1}
\end{equation} 
where (\ref{susysgmcom-2}) holds.
Accordingly $\tilde \lambda_{\mu}(e,\psi)$ 
is determined up to the first order in $\psi$ as follows; 
\begin{equation}
\tilde \lambda_{\mu}(e,\psi) 
= { 1 \over 4i\kappa}( i\kappa^{2}\gamma_{a} \gamma^{\rho} \psi(x) \partial_{[\rho} e{^a}_{\mu]}
- \gamma_{a}\tilde \Lambda{^a}_{\mu} ) = -{i \kappa \over 2}\omega^{ab}{_\mu}(e)\sigma_{ab}\psi,  
\label{lambda-o1}
\end{equation} 
which describes the  Lorentz covariant gravitational spin  coupling of superon. 
Substituting (\ref{lambda-o1}) into (\ref{newsugral-3/2}) 
we obtain the following new LSUSY transformation 
of $\tilde \lambda_{\mu}$(after Fiertz transformations) 
\ba
\delta {\tilde \lambda}_{\mu}(e,\psi) 
\A = \A -{i \kappa \over 2} \{ \delta^{NL}\omega^{ab}{_\mu}(e)\sigma_{ab}\psi + 
\omega^{ab}{_\mu}(e)\sigma_{ab} \delta^{NL}\psi \}  \nonu
\A = \A -{i \over {2 \kappa}} \omega^{ab}{}_\mu(e)\sigma_{ab} \zeta + 
\{ \ O(\psi^{2}) \ + \ O(\psi^{4}) \ \}   \nonu
\A = \A -{i \over {2 \kappa}} \omega^{ab}{}_\mu(e)\sigma_{ab} \zeta + 
\{\tilde\epsilon_{ab}(e,\psi)\sigma^{ab}\cdot \omega^{cd}{}_\mu(e)\sigma_{cd}\psi\cdots \ + \ O(\psi^{4}) \ \}. 
\label{varlambda-o1} 
\ea 
The first term is the intended ordinary global SUSY transformation 
indicating the minimal gravitational interaction of $\tilde \lambda_{\mu}(e,\psi)$ as in SUGRA.  
The second term is the redundant term with higher orders of superon 
and contains the terms recasted as   the Lorentz transformation of  $\tilde \lambda_{\mu}(e,\psi)$ 
with the field dependent parameters. 
(\ref{lambda-o1}) is the SUSY invariant relations for $\tilde \lambda_{\mu}(e,\psi)$, 
for the SUSY transformation of (\ref{lambda-o1}) gives the right hand side of (\ref{newsugral-3/2}) 
with the  appropriate auxiliary fields as shown later. 
Interestingly  the commutator of the two L SUSY transformations  on (\ref{lambda-o1}) induces 
GL(4R); 
\begin{equation}
[\delta_{\zeta_1}, \delta_{\zeta_2}]{\tilde \lambda}_{\mu}(e, \psi)  
= \Xi^{\rho} \partial_{\rho} {\tilde \lambda}_{\mu}(e, \psi)  
+ \partial_{\mu} \Xi^{\rho}{\tilde \lambda}_{\rho}(e, \psi),  
\label{varlambda-o1-comm}
\end{equation}                   
where $\Xi^{\rho}$ is the same field dependent parameter as given in (\ref{susysgmcom-2}).  \\
%
%
%
As for the redundant higher order terms in (\ref{varlambda-o1})  
we can  adjust them by considering the modified spin connection $\tilde \omega^{ab}{_\mu}(e, \psi)$ 
particularly with the contorsion terms and by recasting them in terms of  (the auxiliary field d.o.f.) 
$\tilde \Phi_{\mu}(e,\psi)$. 
In fact,  we found that the following supermultiplet containing 
160 (= 80 bosonic + 80 fermionic) d.o.f. may be 
the supermultiplet of the SUGRA-like LSUSY theory 
which is equivalent to SGM;     \\
for 80 bosonic d.o.f.
\ba 
\A \A
[ \ \tilde e{^a}_{\mu}(e,\psi), a_{\mu}(e,\psi), b_{\mu}(e,\psi), M(e,\psi), N(e,\psi),  \nonumber   \\  
\A \A 
A_{\mu}(e,\psi), B_{\mu}(e,\psi), A{^a}_{\mu}(e,\psi), B{^a}_{\mu}(e,\psi), A{^{[ab]}}_{\mu}(e,\psi) \ ]
\label{80bosons}
\ea
and for 80 fermionic d.o.f.  \\
\ba 
[ \ \tilde{\lambda}{}_{\mu\alpha}(e,\psi), \  \tilde{\Lambda}{^a}_{\mu\alpha}(e,\psi) \ ],
\label{80fermions}
\ea
where $\alpha=1,2,3,4$ are indices for  Majorana spinor. The gauge d.o.f. of 
the local GL(4R) and the local Lorentz of the vierbein are subtracted. 
Note that the second line of (\ref{80bosons}) is equivalent to an auxiliary field with spin 3.    \\
The a priori gauge invariance for  $\tilde \lambda_{\mu\alpha}(e,\psi)$ is not necessary for massive case\cite{f}
corresponding to the spontaneous SUSY breaking. 
For it is natural to suppose that the equivalent linear theory 
may be a coupled system of graviton and massive spin 3/2  with the spontaneous global SUSY breaking, 
which may be an analogue obtained by the super-Higgs mechanism 
in the spontaneous local SUSY breaking of N=1 SUGRA\cite{dz}.   \\
By continuing the heuristic and perturvative arguments 
referring to the familiar SUGRA supermultiplet 
we find the following SUSY invariant relations\cite{sts-1}: 
\ba 
\tilde e{^a}_{\mu}(e,\psi) 
\A=\A 
e{^a}_{\mu}, 
\label{compo-2} \\
%
\tilde \lambda_{\mu}(e,\psi) 
\A=\A 
-i\kappa(\sigma_{ab}\psi)\omega^{ab}{}_{\mu},    
\label{compo-3/2} \\
%
%
\tilde{\Lambda}{}^a{}_{\mu}(e,\psi)
\A=\A
\frac{\kappa^2}{2}\epsilon^{abcd}(\gamma_5\gamma_d\psi)\omega_{bc\mu}, 
\label{compo-5/2} \\
%
A_{\mu}(e,\psi)
\A=\A 
\frac{i\kappa^2}{4}
[(\bar{\psi}\gamma^{\rho}\partial_{\rho}\tilde{\lambda}_{\mu})
-(\bar{\psi}\gamma^{\rho}\tilde{\lambda}_a)\partial_{\mu}e^a{}_{\rho}
-(\bar{\tilde{\lambda}}_{\rho}\gamma^{\rho}\partial_{\mu}\psi)]  \nonumber \\
\A \A
+\frac{\kappa^3}{4}
[(\bar{\psi}\sigma^{a\rho}\gamma^{b}\partial_{\rho}\psi)
(\omega_{\mu ba}+\omega_{ab\mu}) 
+(\bar{\psi}\sigma^{ab}\gamma^{c}\partial_{\mu}\psi)\omega_{cab} ] \nonumber \\
\A \A
+\frac{\kappa^2}{8}(\bar{\tilde{\lambda}}_{\mu}\sigma_{ab}\gamma^{\rho}\psi)\omega^{ab}{}_{\rho},  
\label{compo-Amu} \\
%
B_{\mu}(e,\psi)
\A=\A
\frac{i\kappa^2}{4}
[-(\bar{\psi}\gamma_5\gamma^{\rho}\partial_{\rho}\tilde{\lambda}_{\mu})
+(\bar{\psi}\gamma_5\gamma^{\rho}\tilde{\lambda}_a)\partial_{\mu}e^a{}_{\rho}
-(\bar{\tilde{\lambda}}_{\rho}\gamma_5\gamma^{\rho}\partial_{\mu}\psi)]   \nonumber \\
\A \A
+\frac{\kappa^3}{4}
[(\bar{\psi}\gamma_5\sigma^{a\rho}\gamma^{b}\partial_{\rho}\psi)
(\omega_{\mu ba}+\omega_{ab\mu}) 
+(\gamma_5\sigma^{ab}\gamma^{c}\partial_{\mu}\psi)\omega_{cab} ] \nonumber \\
\A \A
+\frac{\kappa^2}{8}(\bar{\tilde{\lambda}}_{\mu}\gamma_5\sigma_{ab}\gamma^{\rho}\psi)\omega^{ab}{}_{\rho}, 
\label{compo-Bmu} \\
%
A^a{}_{\mu}(e,\psi)
\A=\A
\frac{i\kappa^2}{4}
[(\gamma^{\rho}\gamma^a\partial_{\rho}\tilde{\lambda}_{\mu})
-(\gamma^{\rho}\gamma^a\tilde{\lambda}_b)\partial_{\mu}\tilde{e}{}^b{}_{\rho}
+(\bar{\tilde{\lambda}}_{\rho}\gamma^a\gamma^{\rho}\partial_{\mu}\psi)]   \nonumber \\
\A \A
+\frac{\kappa^3}{4}
[-(\bar{\psi}\sigma^{b\rho}\gamma^a\gamma^{c}\partial_{\rho}\psi)
(\omega_{\mu cb}+\omega_{bc\mu})
-(\gamma^{bc}\sigma^a\gamma^{d}\partial_{\mu}\psi)\omega_{dbc} ]   \nonumber \\
\A \A
-\frac{\kappa^2}{8}(\bar{\tilde{\lambda}}_{\mu}\sigma_{bc}\gamma^a\gamma^{\rho}\psi)\omega^{ab}{}_{\rho}, 
\label{compo-Aamu} \\
%
%
B^a{}_{\mu}(e,\psi)
\A=\A 
\frac{i\kappa^2}{4}
[(\bar{\psi}\gamma_5\gamma^{\rho}\gamma^a\partial_{\rho}\tilde{\lambda}_{\mu})
-(\gamma_5\gamma^{\rho}\gamma^a\tilde{\lambda}_b)\partial_{\mu}\tilde{e}{}^b{}_{\rho}
+({\tilde{\lambda}}_{\rho}\gamma_5\gamma^a\gamma^{\rho}\partial_{\mu}\psi)]    \nonumber \\
\A \A
+\frac{\kappa^3}{8}
[-(\bar{\psi}\gamma_5\sigma^{b\rho}\gamma^a\gamma^{c}\partial_{\rho}\psi)
(\omega_{\mu cb}+\omega_{bc\mu})
-(\bar{\psi}\gamma_5\sigma^{bc}\gamma^a\gamma^{d}\partial_{\mu}\psi)\omega_{dbc} ]   \nonumber \\
\A \A
-\frac{\kappa^2}{8}(\bar{\tilde{\lambda}}_{\mu}\gamma_5\sigma_{bc}\gamma^a\gamma^{\rho}\psi)\omega^{ab}{}_{\rho},  
\label{compo-Bamu} \\
%
A^{[ab]}{}_{\mu}(e,\psi)
\A=\A
\frac{i\kappa^2}{2}
[(\bar{\psi}\gamma^{\rho}\sigma^{ab}\partial_{\rho}\tilde{\lambda}_{\mu})
-(\bar{\psi}\gamma^{\rho}\sigma^{ab}\tilde{\lambda}_c)\partial_{\mu}\tilde{e}{}^c{}_{\rho}
+(\bar{\tilde{\lambda}}_{\rho}\sigma^{ab}\gamma^{\rho}\partial_{\mu}\psi)]       \nonumber \\
\A \A
-\frac{\kappa^3}{2}
[(\bar{\psi}\sigma^{c\rho}\sigma^{ab}\gamma^{d}\partial_{\rho}\psi)
(\omega_{\mu dc}+\omega_{cd\mu})
+(\bar{\psi}\sigma^{cd}\sigma^{ab}\gamma^{e}\partial_{\mu}\psi)\omega_{ecd} ]   \nonumber \\
\A \A
-\frac{\kappa^2}{4}(\bar{\tilde{\lambda}}_{\mu}\sigma_{cd}\sigma^{ab}\gamma^{\rho}\psi)\omega^{ab}{}_{\rho}. 
\label{compo-Aabmu}
\ea
In fact we can show that the following LSUSY transformations on (\ref{80bosons}) and (\ref{80fermions}) 
inuced by  NLSUSY (\ref{newsusy}) close among them(80+80 linearized multiplet) 
at least up to the order with $\psi^{2}$ of superon. 
The contorsion of SUGRA-type breaks the closure and are excluded, so far.
We show  the explicit forms of some of the LSUSY transformations up to  $O(\psi)$. 
\ba
\delta \tilde{e}{}^a{}_{\mu} 
\A=\A
i\kappa \bar{\zeta}\gamma^a\tilde{\lambda}_{\mu}
-\epsilon^a{}_b\tilde{e}{}^b{}_{\mu}
+\bar{\zeta}\tilde{\Lambda}^a{}_{\mu} , 
\label{lsusy-compo-2} \\
\delta \tilde{\lambda}{}_{\mu} 
\A=\A 
-\frac{i}{\kappa}(\sigma_{ab}\zeta)\omega^{ab}{}_{\mu}
+\frac{i}{2}\epsilon^{ab}(\sigma_{ab}\tilde{\lambda}_{\mu})    \nonumber \\
\A \A
+A_{\mu}\zeta +B_{\mu}(\gamma_5\zeta)+A^a{}_{\mu}(\gamma_a\zeta)
+B^a{}_{\mu}(\gamma_5\gamma_a\zeta)+A^{ab}{}_{\mu}(\sigma_{ab}\zeta),
\label{lsusy-compo-3/2} \\
\delta \tilde{\Lambda}{^a}_{\mu}
\A=\A
\frac{1}{2}\epsilon^{abcd}(\gamma_5\gamma_d\zeta)\omega_{bc\mu},
\label{lsusy-compo-5/2} \\
%
\delta A{}_{\mu}
\A=\A
-\frac{1}{8}\left[
	i(\bar{\zeta}\gamma^{\rho}D_{\rho}\tilde{\lambda}_{a})\tilde{e}{}^a{}_{\mu}
	+3i(\bar{\zeta}\gamma^{a}D_{\mu}\tilde{\lambda}_{a})
	+2(\bar{\zeta}\sigma^{\nu\rho}\gamma_{\mu}D_{\nu}\tilde{\lambda}_{\rho})
	\right]
	\nonumber \\
\A \A
-\frac{1}{4\kappa}
	\left[
	3(\bar{\zeta}D_{\mu}\tilde{\Lambda}^a{}_{a})
	+i(\bar{\zeta}\sigma^{ab}D_{\mu }\tilde{\Lambda}_{ab})
	+i(\bar{\zeta}\sigma^{a\rho}D_{\rho }\tilde{\Lambda}_{(ab)})\tilde{e}{}^b{}_{\mu}
	\right]
	\nonumber \\
\A \A
+\frac{1}{16}
	\left[
	4i(\bar{\zeta}\gamma^{\rho}\tilde{\lambda}_a)\omega^a{}_{\rho\mu}
	+4(\bar{\zeta}\sigma^{bc}\gamma^a\tilde{\lambda}{}_{a})\omega_{bc\mu}
	-4(\bar{\zeta}\sigma^{a\rho}\gamma^b\tilde{\lambda}_{[\rho})\omega_{|ab|\mu ]}
	\right.
\nonumber \\
\A \A
	\hspace{1cm}\left.
	+4(\bar{\zeta}\sigma^{ab}\gamma^{c}\tilde{\lambda}_{a})\omega_{\mu cb}
	-3(\bar{\zeta}\sigma^{\rho}\gamma^{bc}\tilde{\lambda}_{[\rho})\omega_{|bc|\mu ]}
	+2i(\bar{\zeta}\sigma^{ab}\gamma_{\mu}\sigma^{cd}\tilde{\lambda}_{a})\omega_{cdb}
	\right]
	\nonumber \\
\A \A
-\frac{1}{8\kappa}
	\left[
	(\bar{\zeta}\gamma^{b}\gamma^a\sigma^{cd}{\tilde{\Lambda}}_{ab})
	+(\bar{\zeta}\sigma^{cd}\gamma^{b}\gamma^a{\tilde{\Lambda}}_{ab})
	\right]\omega_{cd\mu} , 
\label{lsusy-compo-1} \\
%
\delta B{}_{\mu}
\A=\A
-\frac{1}{8}\left[
	5i(\bar{\zeta}\gamma_5\gamma^{\rho}D_{\rho}\tilde{\lambda}_{a})\tilde{e}{}^a{}_{\mu}
	+3i(\bar{\zeta}\gamma_5\gamma^{a}D_{\mu}\tilde{\lambda}_{a})
	+2(\bar{\zeta}\gamma_5\sigma^{\nu\rho}\gamma_{\mu}D_{\nu}\tilde{\lambda}_{\rho})
	\right]
	\nonumber \\
\A \A
-\frac{1}{4\kappa}
	\left[
	3(\bar{\zeta}\gamma_5D_{\mu}\tilde{\Lambda}^a{}_{a})
	+i(\bar{\zeta}\gamma_5\sigma^{ab}D_{\mu }\tilde{\Lambda}_{ab})
	+i(\bar{\zeta}\gamma_5\sigma^{a\rho}D_{\rho }\tilde{\Lambda}_{(ab)})\tilde{e}{}^b{}_{\mu}
	\right]
	\nonumber \\
\A \A
+\frac{1}{16}
	\left[
	-4i(\bar{\zeta}\gamma_5\gamma^{\rho}\tilde{\lambda}_a)\omega^a{}_{\rho\mu}
	+4(\bar{\zeta}\gamma_5\sigma^{cd}\gamma^b\tilde{\lambda}{}_{b})\omega_{cd\mu}
	-4(\bar{\zeta}\gamma_5\sigma^{a\rho}\gamma^b\tilde{\lambda}_{[\rho})\omega_{|ab|\mu ]}
	\right.
\nonumber \\
\A \A
	\hspace{1cm}\left.
	+4(\bar{\zeta}\gamma_5\sigma^{ab}\gamma^{c}\tilde{\lambda}_{a})\omega_{\mu cb}
	-3(\bar{\zeta}\gamma_5\gamma^{\rho}\sigma^{bc}\tilde{\lambda}_{[\rho})\omega_{|bc|\mu ]}
	+2i(\bar{\zeta}\gamma_5\sigma^{ab}\gamma_{\mu}\sigma^{cd}\tilde{\lambda}_{a})\omega_{cdb}
	\right]
	\nonumber \\
\A \A
+\frac{1}{8\kappa}
	\left[
	-(\bar{\zeta}\gamma_5\sigma^{a\rho}\sigma^{bc}\tilde{\Lambda}_{a[\rho})\omega_{|bc|\mu ]}
	+(\bar{\zeta}\gamma_5\sigma^{\nu\rho}\sigma^{ab}\tilde{\Lambda}_{\mu\nu})\omega_{ab\rho}
	-2i(\bar{\zeta}\gamma_5\sigma^{cd}\tilde{\Lambda}^a{}_{a})\omega_{cd\mu}
	\right.
	\nonumber \\
\A \A
	\hspace{1cm}\left.
	+2(\bar{\zeta}\gamma_5\sigma^{ab}\sigma^{cd}{\tilde{\Lambda}}_{ab})\omega_{cd\mu}
	+2(\bar{\zeta}\gamma_5\sigma^{cd}\sigma^{ab}{\tilde{\Lambda}}_{ab})\omega_{cd\mu}
	\right], 
\label{lsusy-compo-2} \\
%
\delta A^a{}_{\mu}
\A=\A
\frac{1}{8}\left[
	-4i(\bar{\zeta}D_{\mu}\tilde{\lambda}^{a})
	+i(\bar{\zeta}\gamma^a\gamma^{\rho}D_{[\mu}\tilde{\lambda}_{\rho ]})
	+2(\bar{\zeta}\sigma^{\nu\rho}\gamma^a\gamma_{\mu}D_{\nu}\tilde{\lambda}_{\rho})
	\right]
	\nonumber \\
\A \A
+\frac{1}{4\kappa}
	\left[
	-i(\bar{\zeta}\sigma^{b\rho}\gamma^aD_{[\mu}\tilde{\Lambda}_{|b|\rho ]})
	-i(\bar{\zeta}\sigma^{\nu\rho}\gamma^aD_{\nu }\tilde{\Lambda}_{b\rho})\tilde{e}{}^b{}_{\mu}
	+(\bar{\zeta}\gamma^c\gamma^b\gamma^aD_{\mu}\tilde{\Lambda}_{bc})
	\right]
	\nonumber \\
\A \A
+\frac{1}{16}
	\left[
	-4i(\bar{\zeta}\gamma^{\rho}\gamma^a\tilde{\lambda}_b)\omega^b{}_{\rho\mu}
	-2(\bar{\zeta}\gamma^{\rho}\gamma^a\sigma^{bc}\tilde{\lambda}_{[\rho})\omega_{|bc|\mu ]}
	+2(\bar{\zeta}\gamma^a\sigma^{cd}\gamma^b\tilde{\lambda}_{b})\omega_{cd\mu}
	\right.
	\nonumber \\
\A \A
	\hspace{1cm}\left.
	+2(\bar{\zeta}\sigma^{cd}\gamma^a\gamma^b\tilde{\lambda}_{b})\omega_{cd\mu}
	+4(\bar{\zeta}\sigma^{b\rho}\gamma^a\gamma^c\tilde{\lambda}_{[\rho})\omega_{|bc|\mu ]}
	-4(\bar{\zeta}\sigma^{bc}\gamma^a\gamma^{d}\tilde{\lambda}_{b})\omega_{\mu dc}
	\right.
	\nonumber \\
\A \A
	\hspace{1cm}\left.
	-(\bar{\zeta}\gamma^a\gamma^{\rho}\sigma^{cd}\tilde{\lambda}_{[\rho})\omega_{|cd|\mu ]}
	-2(\bar{\zeta}\sigma^{bc}\gamma^a\gamma_{\mu}\sigma^{de}\tilde{\lambda}_{b})\omega_{dec}
	\right]
	\nonumber \\
\A \A
+\frac{1}{8\kappa}
	\left[
	(\bar{\zeta}\sigma^{b\rho}\gamma^a\sigma^{cd}\tilde{\Lambda}_{b[\rho})\omega_{|cd|\mu ]}
	-(\bar{\zeta}\sigma^{\nu\rho}\gamma^{a}\sigma^{bc}\tilde{\Lambda}_{\mu\nu})\omega_{bc\rho}
	+i(\bar{\zeta}\gamma^c\gamma^b\gamma^a\sigma^{de}\tilde{\Lambda}_{bc})\omega_{de\mu}
	\right.
	\nonumber \\
\A \A
	\hspace{1cm}\left.
	+i(\bar{\zeta}\sigma^{de}\gamma^c\gamma^b\gamma^a\tilde{\Lambda}_{bc})\omega_{de\mu}
	\right]
	\nonumber \\
\A \A
+\frac{\kappa}{2}(\bar{\zeta}D_{\mu}\Lambda^{\prime}{}^a) 
-\frac{\kappa}{4}(\bar{\zeta}\gamma^c\gamma^a\Lambda^{\prime}{}^b)\omega_{bc\mu}, 
\label{lsusy-compo-3} \\
\delta B^a{}_{\mu}
\A=\A
\frac{1}{8}\left[
	-4i(\bar{\zeta}\gamma_5D_{\mu}\tilde{\lambda}^{a})
	+i(\bar{\zeta}\gamma_5\gamma^a\gamma^{\rho}D_{[\mu}\tilde{\lambda}_{\rho ]})
	+2(\bar{\zeta}\gamma_5\sigma^{\nu\rho}\gamma^a\gamma_{\mu}D_{\nu}\tilde{\lambda}_{\rho})
	\right]
	\nonumber \\
\A \A
+\frac{1}{4\kappa}
	\left[
	-i(\bar{\zeta}\gamma_5\sigma^{b\rho}\gamma^aD_{[\mu}\tilde{\Lambda}_{|b|\rho ]})
	-i(\bar{\zeta}\gamma_5\sigma^{\nu\rho}\gamma^aD_{\nu }\tilde{\Lambda}_{b\rho})\tilde{e}{}^b{}_{\mu}
	+(\bar{\zeta}\gamma_5\gamma^c\gamma^b\gamma^aD_{\mu}\tilde{\Lambda}_{bc})
	\right]
	\nonumber \\
\A \A
+\frac{1}{16}
	\left[
	-4i(\bar{\zeta}\gamma_5\gamma^{\rho}\gamma^a\tilde{\lambda}_b)\omega^b{}_{\rho\mu}
	-2(\bar{\zeta}\gamma_5\gamma^{\rho}\gamma^a\sigma^{bc}\tilde{\lambda}_{[\rho})\omega_{|bc|\mu ]}
	+2(\bar{\zeta}\gamma_5\gamma^a\sigma^{cd}\gamma^b\tilde{\lambda}_{b})\omega_{cd\mu}
	\right.
	\nonumber \\
\A \A
	\hspace{1cm}\left.
	+2(\bar{\zeta}\gamma_5\sigma^{cd}\gamma^a\gamma^b\tilde{\lambda}_{b})\omega_{cd\mu}
	+4(\bar{\zeta}\gamma_5\sigma^{b\rho}\gamma^a\gamma^c\tilde{\lambda}_{[\rho})\omega_{|bc|\mu ]}
	-4(\bar{\zeta}\gamma_5\sigma^{bc}\gamma^a\gamma^{d}\tilde{\lambda}_{b})\omega_{\mu dc}
	\right.
	\nonumber \\
\A \A
	\hspace{1cm}\left.
	-(\bar{\zeta}\gamma_5\gamma^a\gamma^{\rho}\sigma^{cd}\tilde{\lambda}_{[\rho})\omega_{|cd|\mu ]}
	-2(\bar{\zeta}\gamma_5\sigma^{bc}\gamma^a\gamma_{\mu}\sigma^{de}\tilde{\lambda}_{b})\omega_{dec}
	\right]
	\nonumber \\
\A \A
+\frac{1}{8\kappa}
	\left[
	(\bar{\zeta}\gamma_5\sigma^{b\rho}\gamma^a\sigma^{cd}\tilde{\Lambda}_{b[\rho})\omega_{|cd|\mu ]}
	-(\bar{\zeta}\gamma_5\sigma^{\nu\rho}\gamma^{a}\sigma^{bc}\tilde{\Lambda}_{\mu\nu})\omega_{bc\rho}
	+i(\bar{\zeta}\gamma_5\gamma^c\gamma^b\gamma^a\sigma^{de}\tilde{\Lambda}_{bc})\omega_{de\mu}
	\right.
	\nonumber \\
\A \A
	\hspace{1cm}\left.
	+i(\bar{\zeta}\gamma_5\sigma^{de}\gamma^c\gamma^b\gamma^a\tilde{\Lambda}_{bc})\omega_{de\mu}
	\right]
	\nonumber \\
\A \A
+\frac{\kappa}{2}(\bar{\zeta}\gamma_5D_{\mu}\Lambda^{\prime}{}^a) 
-\frac{\kappa}{4}(\bar{\zeta}\gamma_5\gamma^c\gamma^a\Lambda^{\prime}{}^b)\omega_{bc\mu}, 
\label{lsusy-compo-4} \\
%
\delta A^{[ab]}{}_{\mu}
\A=\A
\frac{1}{4}\left[
	-2i(\bar{\zeta}\gamma^{\rho}\sigma^{ab}D_{\rho}\tilde{\lambda}_{c})\tilde{e}{}^c{}_{\mu}
	+i(\bar{\zeta}\sigma^{ab}\gamma^{\rho}D_{\rho}\tilde{\lambda}_{c})\tilde{e}{}^c{}_{\mu}
	+i(\bar{\zeta}\sigma^{ab}\gamma^cD_{\mu}\tilde{\lambda}{}_{c}) 
	-2(\bar{\zeta}\sigma^{\nu\rho}\sigma^{ab}\gamma_{\mu}D_{\nu}\tilde{\lambda}_{\rho})
	\right]
	\nonumber \\
\A \A
+\frac{1}{2\kappa}
	\left[
	-(\bar{\zeta}\sigma^{ab}D_{\mu}\tilde{\Lambda}^{c}{}_{c}) 
	+i(\bar{\zeta}\sigma^{cd}\sigma^{ab}D_{\mu}\tilde{\Lambda}_{cd})
	+i(\bar{\zeta}\sigma^{c\rho}\sigma^{ab}D_{\rho }\tilde{\Lambda}_{(cd)})\tilde{e}{}^d{}_{\mu}
	\right]
	\nonumber \\
\A \A
+\frac{1}{8}
	\left[
	4i(\bar{\zeta}\gamma^{\rho}\sigma^{ab}\tilde{\lambda}_c)\omega^c{}_{\rho\mu}
	+4(\bar{\zeta}\sigma^{c\rho}\sigma^{ab}\gamma^d\tilde{\lambda}_{[\rho})\omega_{|cd|\mu ]}
	-4(\bar{\zeta}\sigma^{cd}\sigma^{ab}\gamma^{e}\tilde{\lambda}_{c})\omega_{\mu ed}
	\right.
\nonumber \\
\A \A
	\hspace{1cm}\left.
	-(\bar{\zeta}\sigma^{ab}\gamma^{\rho}\sigma^{de}\tilde{\lambda}_{[\rho})\omega_{|de|\mu ]}
	-2i(\bar{\zeta}\sigma^{cd}\sigma^{ab}\gamma_{\mu}\sigma^{ef}\tilde{\lambda}_{c})\omega_{efd}
	\right.
	\nonumber \\
\A \A
	\hspace{1cm}\left.
	-4i(\bar{\zeta}\sigma^{cd}\sigma^{ab}\sigma^{ef}\gamma_c\tilde{\lambda}{}_{d})\omega_{ef\mu}
	+2(\bar{\zeta}\sigma^{ef}\sigma^{ab}\gamma^d\tilde{\lambda}{}_{d})\omega_{ef\mu}
	\right]
	\nonumber \\
\A \A
+\frac{1}{4\kappa}
	\left[
	-4(\bar{\zeta}\sigma^{[b|c}\tilde{\Lambda}^{d|}{}_{d})\omega^{a]}{}_{c\mu}
	+i(\bar{\zeta}\sigma^{ab}\sigma^{cd}\tilde{\Lambda}^{e}{}_{e})\omega_{cd\mu}
	-(\bar{\zeta}\sigma^{cd}\sigma^{ab}\sigma^{ef}{\tilde{\Lambda}}_{cd})\omega_{ef\mu}
	\right.
	\nonumber \\
\A \A
	\hspace{1cm}\left.
	-(\bar{\zeta}\sigma^{c\rho}\sigma^{ab}\sigma^{de}\tilde{\Lambda}_{(c\mu )})\omega_{de\rho}
	-2(\bar{\zeta}\sigma^{ef}\sigma^{cd}\sigma^{ab}{\tilde{\Lambda}}_{cd})\omega_{ef\mu}
	\right], 
\label{lsusy-compo-5}
\ea
where $\epsilon^{ab}$ is the Lorentz parameter and we put 
$\epsilon^{ab}=\xi^{\rho}\omega^{ab}{}_{\rho}$. 
Note that the Lorentz transformations are induced on (\ref{lsusy-compo-2}) and (\ref{lsusy-compo-3/2}).
In the right-hand side of (\ref{lsusy-compo-3}) and (\ref{lsusy-compo-4}), 
the last terms contain $\Lambda^{\prime}{}^a{}_{\mu}$ which is defined by 
$\Lambda^{\prime}{}^a{}_{\mu}=-\epsilon^{abcd}\gamma_5\psi\omega_{bcd}$ . 
Note that $\Lambda^{\prime}{}^a{}_{\mu}$ is not the functional of 
the supermultiplet (\ref{80fermions}), 
so we may have to treat $\Lambda^{\prime}{}^a{}_{\mu}$ as new auxiliary field. 
However, if we put $\epsilon^{ab}=\epsilon^{ab}(\tilde{\lambda}{}_{\mu}, \tilde{\Lambda}{}^a{}_{\mu})$, 
e.g. $\epsilon^{ab}=\bar\zeta\gamma^{[a}\tilde{\lambda}{}^{b]}$, 
$\Lambda^{\prime}{}^a{}_{\mu}$ does not appear in 
the right-hand side of (\ref{lsusy-compo-3}) and (\ref{lsusy-compo-4}). 
As a result, the LSUSY transformation 
on the supermultiplet (\ref{80bosons}) and (\ref{80fermions}) 
are written by using the supermultiplet itself 
at least at the leading order of superon $\psi$. 
The higher order terms remain to be studied.    
However we believe that we can obtain the complete linearized off-shell supermultiplets of 
the SP algebra by repeating the similar procedures (on the auxiliary fields) order by order 
which terminates with $(\psi^{4})$.  It may be favorable that  10 bosonic auxiliary fields, for example 
${a_{\mu}(e,\psi), b_{\mu}(e,\psi), M(e,\psi), N(e,\psi)}$ are arbitrary  up now and available 
for the closure of the off-shell SP algebra in all orders.     \par
We show  some general properties of the new NLSUSY algebra and 
discuss  some systematics of the linearization in the next section, which is complementary
for linearizing SGM.      \par
${\bf 3.2 \ Case \ \tilde e{^a}_{\mu}(e, \psi) =  e{^a}_{\mu} 
+ f{^a}_{\mu}[O(\psi^2), ...] }$   \\
In the previous section the linearization has been carried out consistently 
at least in the lowest order 
of the superon field under the simplest SUSY invariant relation for graviton 
$\tilde e{^a}_{\mu}(e, \psi) =  e{^a}_{\mu}$ of (\ref{compo-2}). 
In this section we consider the generalization of (\ref{compo-2}) and for a comparison 
take another way of thinking. 
We adopt the following assumption in stead of (\ref{compo-2}) 
\ba
\tilde e{^a}_{\mu}(e, \psi) =  e{^a}_{\mu} + f{^a}_{\mu}[O(\psi^{2}), ...]. 
\label{new-e}
\ea                               
This means that the vierbein of LSUSY, 
i.e. the asymptotic (low energy) gravitational field,  
has the contribution from the superon-antisuperon (vacuum) 
higher order components.   \\
Accordingly, the variation of $\tilde e{^a}_{\mu}$ 
by means of NLSUSY transformations of $(e{^a}_{\mu}, \psi)$ becomes 
\ba
\delta \tilde e{^a}_\mu = \delta e{^a}_\mu + \delta f{^a}_\mu[O(\psi^1), ...]. 
\label{vari-new-e}
\ea
Substituting 
$\delta \tilde e{^a}_\mu 
= i \kappa \bar\zeta \gamma^a \tilde \lambda_\mu 
+ \bar\zeta \tilde \Lambda{^a}_\mu - \epsilon{^a}_b \tilde e{^b}_\mu$ 
\ ($\epsilon{^a}_b = \bar\zeta \Gamma{^a}_b \psi$, 
\ $\Gamma_{ab} = \Gamma_{[ab]}$) and  
$\delta e{^a}_\mu = \xi^\rho \partial_{[\rho} e{^a}_{\mu]}$ 
\ ($\xi^\mu = i \kappa^2 \bar\zeta \gamma^\mu \psi$) 
into Eq.(\ref{vari-new-e}) produces 
\ba
i \kappa \bar\zeta \gamma^a \tilde \lambda_\mu 
+ \bar\zeta (\tilde \Lambda{^a}_\mu - \Gamma{^a}_b \psi \tilde e{^b}_\mu) 
= i \kappa^2 \bar\zeta \gamma^\rho \psi \partial_{[\rho} e{^a}_{\mu]} 
+ \delta f{^a}_\mu[O(\psi^1), ...]. 
\label{all}
\ea
The $\tilde \lambda_\mu$, $\tilde e{^a}_\mu$ (namely, $f{^a}_\mu$), 
$\tilde \Lambda{^a}_\mu$ and $\epsilon{^a}_b$ (namely, $\Gamma{^a}_b$) 
are expanded in terms of $(e{^a}_{\mu}, \psi)$ 
as they satisfy Eq.(\ref{all}) for  all orders of $\psi$.      \\
For example, we have from Eq.(\ref{all}) for the terms with $O(\psi^1)$, 
\ba
i \kappa \bar\zeta \gamma^a \tilde \lambda_\mu[O(\psi^1)] 
+ \bar\zeta (\tilde \Lambda{^a}_\mu[O(\psi^1)] 
- \Gamma{^a}_b[O(\psi^0)] \psi e{^b}_\mu) \nonu
= i \kappa^2 \bar\zeta \gamma^\rho \psi \partial_{[\rho} e{^a}_{\mu]}
+ \delta f{^a}_\mu[O(\psi^1)]. 
\label{first-o}
\ea
Let us consider the example of $\tilde \lambda_\mu[O(\psi^1)] 
= - i \kappa \omega^{ab}{_\mu}(e) \sigma_{ab} \psi$ \ following (\ref{lambda-o1}) 
and Lorentz parameter $\epsilon{^a}_b =  \ \xi^\rho \omega_{\rho}{^a}{_b}(e)$, 
i.e., $\Gamma{^a}_b =  \ i \kappa^2 \gamma^\rho \omega{^a}_{b \rho}(e)$. 
When we substitute them into (\ref{first-o}), we have 
the relation to decide the form of $\tilde \Lambda{^a}_\mu[O(\psi^1)]$ 
and $\delta f{^a}_\mu[O(\psi^1)]$ as 
\ba
- {1 \over 2} \kappa^2 \epsilon{^a}_{bcd} 
\bar\zeta \gamma_5 \gamma^{b} \psi \omega{^{cd}}_\mu(e) 
+ \bar\zeta \tilde \Lambda{^a}_\mu[O(\psi^1)] 
= \delta f{^a}_\mu[O(\psi^1)]. 
\label{example}
\ea
If we take $\delta f{^a}_\mu[O(\psi^1)] = 0$ in Eq.(\ref{example}), we have 
the $\tilde e{^a}_\mu = e{^a}_\mu$ case 
with $\tilde \Lambda{^a}_\mu[O(\psi^1)] = (1/2) \kappa^2 \epsilon{^a}_{bcd} 
\bar\zeta \gamma_5 \gamma^{b} \psi \omega{^{cd}}_\mu(e)$ 
as we have already discussed in the previous section 3.1. 
On the other hand, if we put $\tilde \Lambda{^a}_\mu[O(\psi^1)] = 0$ 
in $\delta \tilde e{^a}_\mu$, we obtain 
\ba
f{^a}_\mu[O(\psi^2)] = - {1 \over 4} \kappa^4 \epsilon{^a}_{bcd} 
\bar\psi \gamma_5 \gamma^{b} \psi \omega{^{cd}}_\mu(e). 
\label{f-a-mu}
\ea
Here we note that since the commutator of two NLSUSY transformations 
for (\ref{f-a-mu}) becomes 
\ba
[\delta_1, \delta_2] f{^a}_\mu 
= \Xi^\rho \partial_\rho f{^a}_\mu + \partial_\mu \Xi^\rho f{^a}_\rho 
= \delta_{{\rm GL(4R)}} f{^a}_\mu,
\ea
the commutator of two NLSUSY transformations for $\tilde e{^a}_\mu$ 
also closes on GL(4R) as 
\ba
[\delta_1, \delta_2] \tilde e{^a}_\mu 
= \Xi^\rho \partial_\rho \tilde e{^a}_\mu 
+ \partial_\mu \Xi^\rho \tilde e{^a}_\rho 
= \delta_{{\rm GL(4R)}} \tilde e{^a}_\mu.
\ea
The 64 bosonic auxiliary fields $\tilde \Phi_\mu$ 
(or $A_\mu, B_\mu, A{^a}_\mu, B{^a}_\mu, A{^{[ab]}}_\mu$) 
at the lowest order of $\psi$ are read from 
$\delta \tilde \lambda_\mu[O(\psi^0), O(\psi^2)] 
= - (i/\kappa) \omega{^{ab}}_\mu(e) \sigma_{ab} \zeta 
+ \tilde \Phi_\mu[O(\psi^2)] \zeta$ 
in Eq.(20) up to the contribution of the Lorentz transformation 
$-(i/2) \epsilon^{ab} \sigma_{ab} \tilde \lambda_\mu$ 
as follows; 
\ba
\tilde \Phi_\mu[O(\psi^2)] 
= \A \A - \kappa^3 
[- \sigma_{ab} \partial_\nu \psi \ \bar\psi \gamma^\nu 
e^{a \rho} \partial_{[\rho} e{^b}_{\mu]} 
+ \sigma_{ab} \psi \ \bar\psi \gamma^\sigma 
e^{a \nu} e{_c}^\rho \partial_{[\sigma} e{^c}_{\nu]} 
\partial_{[\rho} e{^b}_{\mu]} \nonu
\A \A - \sigma_{ab} \psi \ e^{a \rho} 
\{ \partial_\rho (\bar\psi \gamma^\nu \partial_{[\nu} e{^b}_{\mu]}) 
- \partial_\mu (\bar\psi \gamma^\nu \partial_{[\nu} e{^b}_{\rho]}) \} \nonu
\A \A - \sigma^{\rho \sigma} \partial_\nu \psi \ \bar\psi \gamma^\nu 
\partial_{\rho} e{^c}_\sigma e_{c \mu} 
+ \sigma^{\nu \sigma} \psi \ \bar\psi \gamma^\lambda 
e{_d}^\rho \partial_{[\lambda} e{^d}_{\nu]} 
\partial_{[\rho} e{^c}_{\sigma]} e_{c \mu} \nonu
\A \A - \sigma^{\rho \sigma} \psi 
\{ \partial_\rho (\bar\psi \gamma^\nu \partial_{[\nu} e{^c}_{\sigma]}) e_{c \mu} 
+ \bar\psi \gamma^\nu \partial_{\rho} e{^c}_\sigma 
\partial_{[\nu} e_{\mid c \mid \mu]}) \} ]. 
\label{b-aux}
\ea
The 64 components of $\tilde \Phi_\mu[O(\psi^2)]$, e.g., 
$A_\mu, B_\mu, A{^a}_\mu, B{^a}_\mu, A{^{[ab]}}_\mu$ 
can be obtained by using Fierz transformations. 
If we define $\hat \Lambda{^a}_\mu[O(\psi^1)] 
= \gamma^\rho \psi \partial_{[\rho} e{^a}_{\mu]} 
= \gamma^\rho \psi \omega{^a}_{[\rho \mu]}$ as 64 fermionic auxiliary fields 
($\tilde \Lambda{^a}_\mu[O(\psi^1)]  = 0$ in $\delta \tilde e{^a}_\mu$ 
in the example now we consider), 
then the the variation of Eq.(\ref{b-aux}) 
by means of NLSUSY transformations of $(e{^a}_{\mu}, \psi)$ 
at the lowest order of $\psi$ 
is written in terms of the fields of the linear supermultiplet, 
$\tilde \lambda_\mu[O(\psi^1)]$ and $\hat \Lambda{^a}_\mu[O(\psi^1)]$; 
namely, $\delta \tilde \Phi_\mu[O(\psi^1)]$ becomes 
\ba
\delta \tilde \Phi_\mu[O(\psi^1)] 
= \A \A - \kappa 
\left[ \ {i \over \kappa} ( \partial_\nu \tilde \lambda_\mu 
\ \bar\zeta \gamma^\nu 
+ \tilde \lambda_\nu 
\ \bar\zeta \partial_\mu e{_c}^\nu \gamma^c ) \right. \nonu
\A \A + \sigma_{ab} \zeta \ \bar{\hat \Lambda}{^c}_\nu 
e^{a \nu} e{_c}^\rho \partial_{[\rho} e{^b}_{\mu]} 
- \sigma_{ab} \zeta \ e^{a \rho} 
( \partial_\rho \bar{\hat \Lambda}{^b}_\mu 
- \partial_\mu \bar{\hat \Lambda}{^b}_\rho ) 
\nonu
\A \A + \sigma^{\nu \sigma} \zeta \ \bar{\hat \Lambda}{^d}_\nu 
e{_d}^\rho \partial_{[\rho} e{^c}_{\sigma]} e_{c \mu} 
\nonu
\A \A - \sigma^{\rho \sigma} \zeta 
( \partial_\rho 
\bar{\hat \Lambda}{^a}_\sigma e_{a \mu} 
+ \bar{\hat \Lambda}_{a \mu} \partial_{\rho} e{^a}_\sigma ) \Bigg] [O(\psi^1)]. 
\ea
The systematic arguments with the generalized assumption for graviton 
(\ref{new-e}) can be continued in principle to higher order terms. 
And it allows more varieties of the way of constructing the SUSY invariant relations 
choosing the auxiliary fields at least at the lowest order of $\psi$.  \par
Finally, we discuss the commutators for more general cases.   \\
Here we consider a functional of $(e{^a}_\mu, \psi)$ and their derivatives as 
\ba
f_A(\psi, \bar\psi, e{^a}_\rho; 
\psi_{,\rho}, \bar\psi_{,\rho}, e{^a}_{\rho,\sigma}),  \ (A = \mu, \mu \nu, ... etc.) 
\ea
with $\psi_{,\rho} = \partial_\rho \psi, etc.$, 
and we suppose that $f_A$ is the functional of $O(\psi^2)$ for simplicity. 
Then we have the variation of $f_A$, 
\ba
\delta f_A = {{\partial f_A} \over {\partial \psi}} \delta \psi 
+ \delta \bar\psi {{\partial f_A} \over {\partial \bar\psi}} 
+ {{\partial f_A} \over {\partial e{^a}_\rho}} \delta e{^a}_\rho 
+ {{\partial f_A} \over {\partial \psi_{,\rho}}} (\delta \psi)_{,\rho} 
+ (\delta \bar\psi)_{,\rho} {{\partial f_A} \over {\partial \bar\psi_{,\rho}}} 
+ {{\partial f_A} \over {\partial e{^a}_{\rho,\sigma}}} 
(\delta e{^a}_\rho)_{,\sigma}.  
\ea
and the commutator for $f_A$ becomes 
\ba
[\delta_1, \delta_2] f_A 
= \A \A {{\partial f_A} \over {\partial \psi}} [\delta_1, \delta_2] \psi 
+ [\delta_1, \delta_2] \bar\psi {{\partial f_A} \over {\partial \bar\psi}} 
+ {{\partial f_A} \over {\partial e{^a}_\rho}} [\delta_1, \delta_2] e{^a}_\rho 
\nonu
\A \A + {{\partial f_A} \over {\partial \psi_{,\rho}}} 
([\delta_1, \delta_2] \psi)_{,\rho} 
+ ([\delta_1, \delta_2] \bar\psi)_{,\rho} 
{{\partial f_A} \over {\partial \bar\psi_{,\rho}}} 
+ {{\partial f_A} \over {\partial e{^a}_{\rho,\sigma}}} 
([\delta_1, \delta_2] e{^a}_\rho)_{,\sigma} 
\label{com-f}
\ea
If we substitute the commutators for $(e{^a}_\mu, \psi)$ of Eq.(8) 
into Eq.(\ref{com-f}), we obtain 
\begin{equation}
[\delta_1, \delta_2] f_A = \Xi^\lambda \partial_\lambda f_A + G_A, 
\label{com2-f}
\end{equation}
where $G_A$ is defined by 
\ba
G_A = \A \A \partial_\rho \Xi^\lambda 
\left( {{\partial f_A} \over {\partial e{^a}_\rho}} e{^a}_\lambda 
+ {{\partial f_A} \over {\partial \psi_{,\rho}}} \partial_\lambda \psi 
+ \partial_\lambda \bar\psi 
{{\partial f_A} \over {\partial \bar\psi_{,\rho}}} 
+ {{\partial f_A} \over {\partial e{^a}_{\sigma,\rho}}} 
\partial_\lambda e{^a}_\sigma 
+ {{\partial f_A} \over {\partial e{^a}_{\rho,\sigma}}} 
\partial_\sigma e{^a}_\lambda \right) 
\nonu
\A \A + \partial_\rho \partial_\sigma \Xi^\lambda 
{{\partial f_A} \over {\partial e{^a}_{\rho,\sigma}}} e{^a}_\lambda. 
\label{GA}
\ea
The first term in r.h.s. of Eq.(\ref{com2-f}) means the translation of $f_A$. 
Therefore Eq.(\ref{com2-f}) shows that the closure of the commutator algebra 
on GL(4R) for the various functionals $f_A$ in the previous argument 
depends on $G_A$ of Eq.(\ref{GA}), 
and these argument reproduces all the previous commutators respectively.

\newsection{Conclusions}

\noindent
Now we summarize the results as follows. Referring to SUGRA transformations on the off-shell 
SUGRA supermultiplet, particulaly to the Lorentz 
transformation, we have obtained the SUSY invariant relations and carried out  the linearization  
explicitly up to $O(\psi^2)$ in the (SUGRA-like) LSUSY transformations. 
We presented two different ways of the linearization 
as the subsection 3.1 and 3.2, but we think that they are complementary 
for finding the correct way to the linearization of higher order terms. 
The d.o.f. of the high spin (auxiliary) fields $A^{ab}{}_{\mu}$(spin 3) and 
$\tilde{\Lambda}{^a}_{\mu\alpha}(e,\psi)$(spin 5/2), though they appear through the arguments of 
Lorentz transformation, may reflect the characteristic structure of the tangent space  
of SGM spacetime, which is unstable. 
Interestingly the linearization mimicking SUGRA seems excluding the naive 
composite picture $\tilde\lambda_{\mu}=\gamma_{\mu} \psi+ \cdots$, which may be suggested from 
the flat space linearization. 
The LSUSY transformations on the two different types of the linearized 
supermultiplet are different from SUGRA but close on the algebra 
isomorophic to SP up to  $O(\psi^{2})$ 
in the SUSY invariant relations. 
The complete linearization in all orders, 
which can be anticipated by the systematics emerging in the present study, needs specifications of 
the auxiliary fields and remains to be studied.  The subsequent construction of the invariant linear 
SUSY action is challenging.   \\
The linearization of the NLSUSY E-H type SGM action (\ref{SGM}) with the extra dimensions 
gives another unification framework describing the observed particles not only as composites 
but also as elementary fields. 
The systematic linearization by using the superfield formalism 
applied to the coupled system of V-A action with SUGRA
\cite{lr}-\cite{wb} is open 
but may be inevitable to complete the linearization, especially for $ N > 1$. 
The linearization of SGM action for spin 3/2 N-G fermion field\cite{st3} 
(with extra dimensions) 
may be in the same scope and gives the deep insight into the structure of SGM.  \\

\vskip 15mm

One(K.S.) of the authors would like to thank J. Wess for useful and enjoyable discussions 
on the algebra and the superfield in Izu. 
They would like to thank U. Lindstr\"om for the interest in our works and for bringing the 
useful references to our attentions. 
The work of M. Sawaguchi is supported in part by the research project of  High-Tech Research Center 
of Saitama Institute of Technology.

\newpage

%
\newcommand{\NP}[1]{{\it Nucl.\ Phys.\ }{\bf #1}}
\newcommand{\PL}[1]{{\it Phys.\ Lett.\ }{\bf #1}}
\newcommand{\CMP}[1]{{\it Commun.\ Math.\ Phys.\ }{\bf #1}}
\newcommand{\MPL}[1]{{\it Mod.\ Phys.\ Lett.\ }{\bf #1}}
\newcommand{\IJMP}[1]{{\it Int.\ J. Mod.\ Phys.\ }{\bf #1}}
\newcommand{\PR}[1]{{\it Phys.\ Rev.\ }{\bf #1}}
\newcommand{\PRL}[1]{{\it Phys.\ Rev.\ Lett.\ }{\bf #1}}
\newcommand{\PTP}[1]{{\it Prog.\ Theor.\ Phys.\ }{\bf #1}}
\newcommand{\PTPS}[1]{{\it Prog.\ Theor.\ Phys.\ Suppl.\ }{\bf #1}}
\newcommand{\AP}[1]{{\it Ann.\ Phys.\ }{\bf #1}}

\end{document}